\documentclass[12pt,a4paper]{article}
\usepackage{amsmath}
\usepackage{amsfonts,graphicx}
\usepackage{amssymb}
\usepackage[left=2cm,right=2cm,top=2cm,bottom=2cm]{geometry}
\usepackage{bm}
\usepackage{physics}
\usepackage[ruled,vlined]{algorithm2e}

\begin{document}

\title{Machine Learning in Nonlinear Dynamical Systems}

\author{Sayan Roy\footnote{sayanr16@iiserb.ac.in} \\ \textit{Department of Physics,}\\ \textit{Indian Institute of Science Education and Research Bhopal,} \\
\textit{Bhopal Bypass Road, Bhauri, Bhopal, Madhya Pradesh, 462066, India} \\ \\ Debanjan Rana\footnote{debanjan16@iiserb.ac.in}  \\ \textit{Department of Chemistry,}\\ \textit{Indian Institute of Science Education and Research Bhopal,}\\
\textit{Bhopal Bypass Road, Bhauri, Bhopal, Madhya Pradesh, 462066, India}} 

\date{}

\maketitle


\begin{abstract}
In this article, we discuss some of the recent developments in applying
machine learning (ML) techniques to nonlinear dynamical
systems. In particular, we demonstrate how to build a suitable ML
framework for addressing two specific objectives of relevance: prediction of future
evolution of a system and unveiling from given time-series data the
analytical form of the underlying dynamics. This article is written in a
pedagogical style appropriate for a course in nonlinear dynamics or
machine learning.
\end{abstract}


\section{Introduction}

Study of dynamics has fascinated mankind for many centuries. One of the
early things that intrigued the human mind was the motion of objects,
both animate and inanimate \cite{CM}. Interest and curiosity in
understanding the motion of planetary objects and various natural
phenomena such as wind, rain, etc. led to the development of the field of nonlinear dynamics as a major branch of study in physics and mathematics, with potential applications in different branches of science and engineering \cite{SH}.  

Nonlinear dynamics has played a crucial role in understanding complex
systems. One is routinely confronted with new phenomena warranting suitable dynamical modelling. In many cases, however, even if such a modelling may be achieved, exact closed-form solutions of the dynamics remain elusive with existing mathematical techniques. 

With vast amount of data being generated and enhanced data management
systems in the world of today, we have tons of data to analyze and
exploit to our advantage.
In this regard, developing a method that may predict future events by
learning from past data would evidently be considered a major
advancement. 
For example, making reliable predictions for the closing price of stock
each day, the issue of occurrence of cardiac arrhythmia from existing
ECG data, the long-term dynamical state for a chaotic system would be
highly desirable. Machine learning algorithms fundamentally work on a
similar strategy of learning from given data, and have proven to be very
efficient in finding patterns from higher-dimensional data such as those
involving images, speech, etc. \cite{DLB}. In 1998, Tom Mitchell in his
textbook \cite{TM} gave a very logical definition of machine learning,
defining it as a Well-posed Learning Problem. He writes, ``\emph{A computer program is said to learn from experience} E \emph{with respect to some class of tasks} T \emph{and performance measure} P \emph{if its performance at tasks in} T, \emph{as measured by} P, \emph{improves with experience} E ". By following this
definition, we propose in this contribution an ML framework that may
predict future data by learning from existing data. For our model, E is
the given time-series data that we feed into the algorithm, T is the task of
prediction, and P  measures the performance of whether it can predict correctly. We will present in this work an explicit application of this technique
in understanding representative dynamical systems for which analytical closed-form  solution is not available.

Another direction that fascinates us about machine learning is the associated data-driven discovery of the governing dynamical equations. Traditionally, nonlinear dynamics has progressed  by invoking fundamental principles and intuitions in developing theoretical explanations of observations. In the present big data era, a central challenge is to reconstruct the underlying dynamical system from an analysis of the existing data. In this backdrop, we discuss here a novel technique called Sparse Identification of Nonlinear Dynamical Systems (SINDy) \cite{SINDy}, developed by \textit{Brunton et al.} based on an ML concept called sparse regression, which is used to unveil governing dynamical laws from time-series data. Underlying the technique is the reasonable assumption 
that most physical systems have simple dynamics containing only a few of the many nonlinear functions possible, so that the governing equation becomes sparse in the high dimensional space of nonlinear functions. We will demonstrate here an application of the algorithm in the context of two paradigmatic nonlinear oscillators, showing in particular how it captures in a very efficient manner the rich nonlinear behavior of the two systems. 

This article is laid out as follows: In Section 2, we describe the
machine learning framework that predicts future data by learning from
existing data, which we then apply to two different time-series data. In
Section 3, we introduce the SINDy algorithm followed by its application
to two paradigmatic nonlinear oscillator systems, namely, the Duffing-Van der Pol oscillator and the R\" ossler attractor. The article ends with key concluding remarks in the last section, followed by an appendix that contains some technical details.     


\section{Prediction with Neural Networks}
In order to start our discussion, let us consider a time-series data of
the form $x(t_{1}), x(t_{2}),\ldots, x(t_{n})$, where $x(t_{i})$ represents
the value of the dynamical variable $x$ at $i$-th  time
instant $t_{i}$ and $n$ is the total number of data points in the
time-series data. We first split the dataset into a training set (70 \%
of total data) and a test set (30 \% of total data). The ML model is
trained in the training set, and then we evaluate the goodness of the
model by comparing its predictions in the test set with the original
test set data. The time-series dataset is restructured in the form shown
in Table \ref{T1}. 

\begin{table}
\begin{center}
\begin{tabular}{ |c|c|c| } 
  \hline
 Input Data & Output Data \\ [0.5ex]
 \hline
$x(t_1),~x(t_2),~x(t_3),~x(t_4),\ldots,~x(t_{20})$   &  $x(t_{21})$ \\ 
$x(t_2),~x(t_3),~x(t_4),~x(t_5),\ldots,~x(t_{21})$   &  $x(t_{22})$ \\ 
$x(t_3),~x(t_4),~x(t_5),~x(t_6),\ldots,~x(t_{22})$   &  $x(t_{23})$ \\ 
$x(t_4),~x(t_5),~x(t_6),~x(t_7),\ldots,~x(t_{23})$   &  $x(t_{24})$ \\ 
$x(t_5),~x(t_6),~x(t_7),~x(t_8),\ldots,x(t_{24})$   &  $x(t_{25})$ \\ 
 \hline
\end{tabular}
\end{center}
\caption{ Restructuring the dataset for neural networks. We choose 20 as the length of the sliding window.Here, the number of data points $N$ used is 6880. A glimpse of the dataset containing 5 elements is shown here. }\label{T1}
\end{table}

This type of restructuring makes it evident that the
output data at a given time is determined in terms of input data from
all previous times over a given time interval (the so-called sliding
time window). This type of ML setting is known as \textbf{supervised learning}.
Mathematically, we define supervised learning as follows. Let  $S \equiv
(x_{1}, y_{1}), (x_{2}, y_{2}),(x_{3}, y_{3},\ldots,(x_{N}, y_{N})$ be the
dataset, where $N$ is the number of data points in this restructured
format, $x_{i}$'s define the input dataset $X$ and $y_{i}$'s constitute
the output dataset $Y$. When $Y$ is continuous, one talks
of a regression problem, while for discrete $Y$, one has a classification problem.
 For example, referring to Table \ref{T1}, we have $x_{1} \equiv
 \{x(t_1), x(t_2),..., x(t_{20})\}$ and $y_{1} \equiv x(t_{21})$. Now,
 there exists a function  $F:X \rightarrow Y$ which satisfies all the
 data points in $S$ but its analytical form is unknown. Machine learning
 aims to identify this mapping. To this end, we choose an ML framework,
 and by using a suitable learning algorithm, we check with the help of
 an accuracy metric whether the framework can approximate well the true
 function. The ML framework and the learning algorithm together form the
 \textit{ML model}. In our setting, the ML framework is a neural
 network, while the learning algorithm is that of gradient descent.\\
 
\textit{Neural Network Architecture}\\

\begin{figure}[h!]
\centering
\includegraphics[width=10 cm, height=6cm]{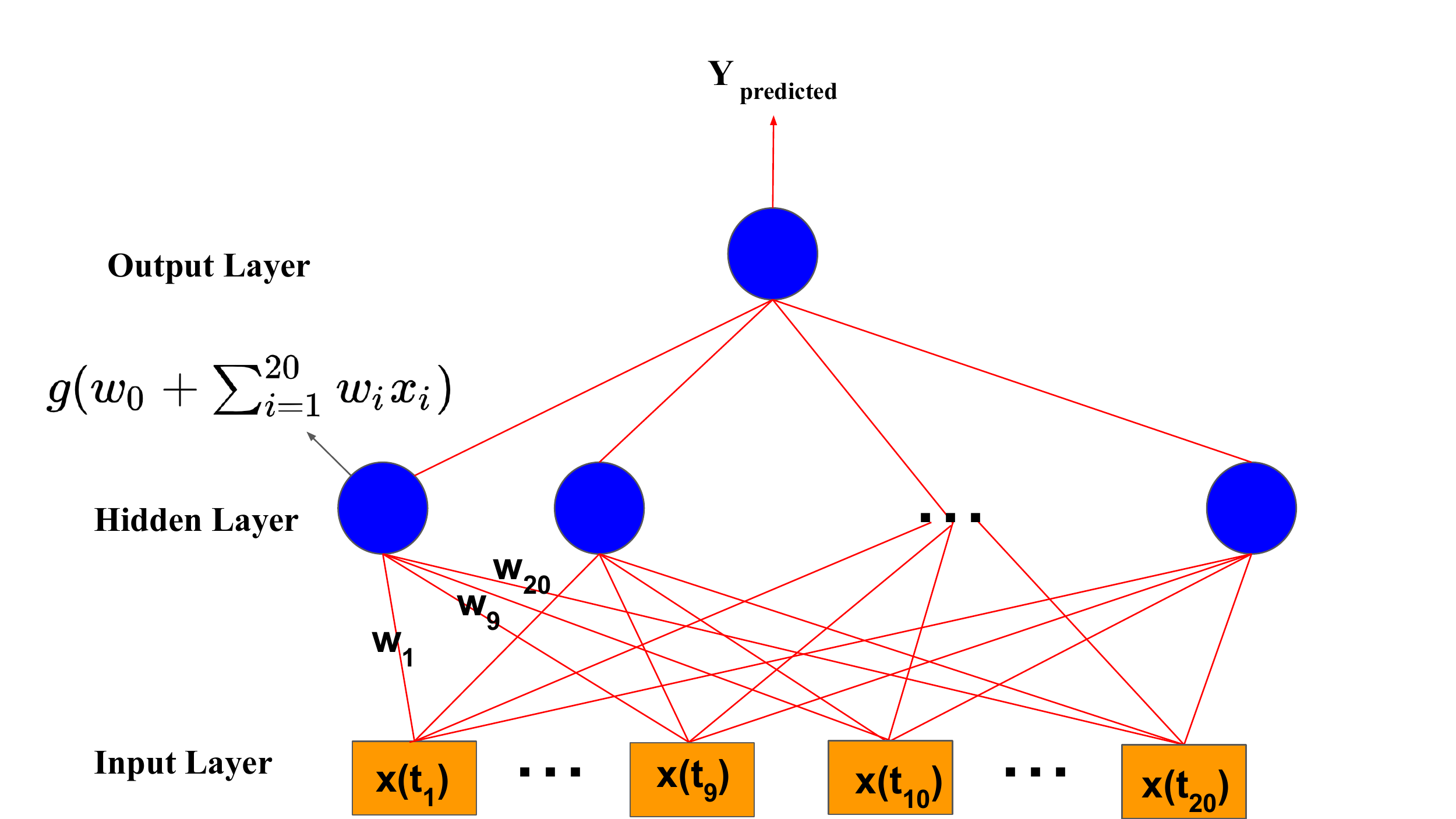}
\caption{Neural Network Architecture consisting of three types of
layers: an input layer, several hidden layers, and an output layer. The
boxes and the circles represent the nodes of the network, while a red
line joining two nodes represents the weight parameter for the corresponding connection and $g$ the activation function. For more details, see text. }\label{Fig-1}
\end{figure}

The concept of neural networks (NNs) is inspired from biological
neurons. A visual representation of an NN is shown in Fig. \ref{Fig-1}.
The first layer is the input layer, and each orange box contains an
input data point. The last layer is called the output layer, while all
the intermediate layers are called the hidden layers. The hidden layers
are constituted by nodes, shown by blue circles in  Fig. \ref{Fig-1},
which are also termed as artificial neurons/ perceptrons. The nodes are
the building blocks of the neural network architecture. Each node in the
first hidden layer receives a set of inputs $\{x(t_{1}), x(t_{2}),...,
x(t_d) \}$ (with $d$ being the length of the sliding time window) from
the previous layer, and these are multiplied with their corresponding weights $\{w_1, w_{2},..., w_d \}$ and summed up. A bias term $w_0$ is added to the sum, which is then passed through an activation function to get the output $\hat{y}$ of that node. The bias term acts as an additional parameter that helps to adjust the output and adds flexibility to the learning process. The purpose of the activation function is to introduce non-linearity into the model, thereby allowing modeling of the output as a nonlinear function of the input data. We thus have

\begin{equation}
\hat{y} = g \left( w_{0} + \sum_{i = 1}^{d}w_{i} x_{i} \right),
\label{Eq:1}
\end{equation}

where $g$ is the activation function. The output from each node is then
passed on as the input to the next layers in the forward direction until
one reaches output layer. In this way, by forming a network of
perceptrons, a neural network gets constructed. The weights and biases
for all the nodes taken together are referred to as the \textit{parameters} $(W)$ of the NN model. Parameters other than weights and biases are termed as \textit{hyperparameters}, e.g. the number of hidden layers, number of neurons (nodes) in a layer, learning rate, and many more \cite{DLB}. After building this NN model, we construct a loss function $L(W)$, as 

\begin{equation}
L(W) \equiv \frac{1}{N}\sum_{i = 1}^{N} (y_{i} - f(x_{i},W))^{2},
\label{Eq:2}
\end{equation}
where $y_{i}$ is the actual output data, $f(x_{i},W)$ is the predicted output with $x_{i}$ representing the input data, and  $N$ is the number of input data points. The physical significance of loss function is that it is a measure of the prediction error of the ML framework.  Now, we learn the parameters $W^{*}$ such that it satisfies
\begin{equation}
W^{*} = \min_{W} L(W).
\label{Eq:3}
\end{equation}

This optimization may be done by using a gradient descent algorithm discussed below.\\ 

\begin{algorithm}[H]
\SetAlgoLined
\textbf{Initialize weights randomly}, W $\sim$ $\mathbb{N}(0,\sigma^2)$ \\
\While{Until convergence i.e., Loss Function $ \geq \epsilon$}{
    \textbf{compute gradient}, $\frac{\partial L(W)}{\partial W}$ \\
    \textbf{update weights},$ W \rightarrow  W +  \eta\frac{\partial
    L(W)}{\partial W}  $} 
\textbf{return} W
 \caption{Gradient Descent Algorithm}
\end{algorithm}
\vspace{0.5cm}
In each iteration, the algorithm computes the gradient of the loss
function with respect to all the parameters $\partial L(W)/\partial W$
by a method called \textit{backpropagation} \cite{DLB}. The parameters
are then updated with their respective gradients. This is how learning
proceeds in an NN model. Once the learning is achieved, if we feed the test data into the NN model, it should give the expected output with good accuracy. The parameter $\eta$, called the
\textit{learning rate}, is responsible for the convergence of the algorithm.

\subsection{Results and Discussions}

Root Mean Square Error (RMSE) was selected as the metric to evaluate the performance of the neural network model. It is defined as
\begin{equation}
\mathrm{RMSE} \equiv \sqrt{\frac{\sum_{i=1}^{N}(y_{i}^{\mathrm{pred}} - y_{i}^{\mathrm{actual}})^2}{N}}.
\label{Eq:4}
\end{equation}

\begin{figure}[h!]

\centering
\includegraphics[width=10 cm, height=5.5cm]{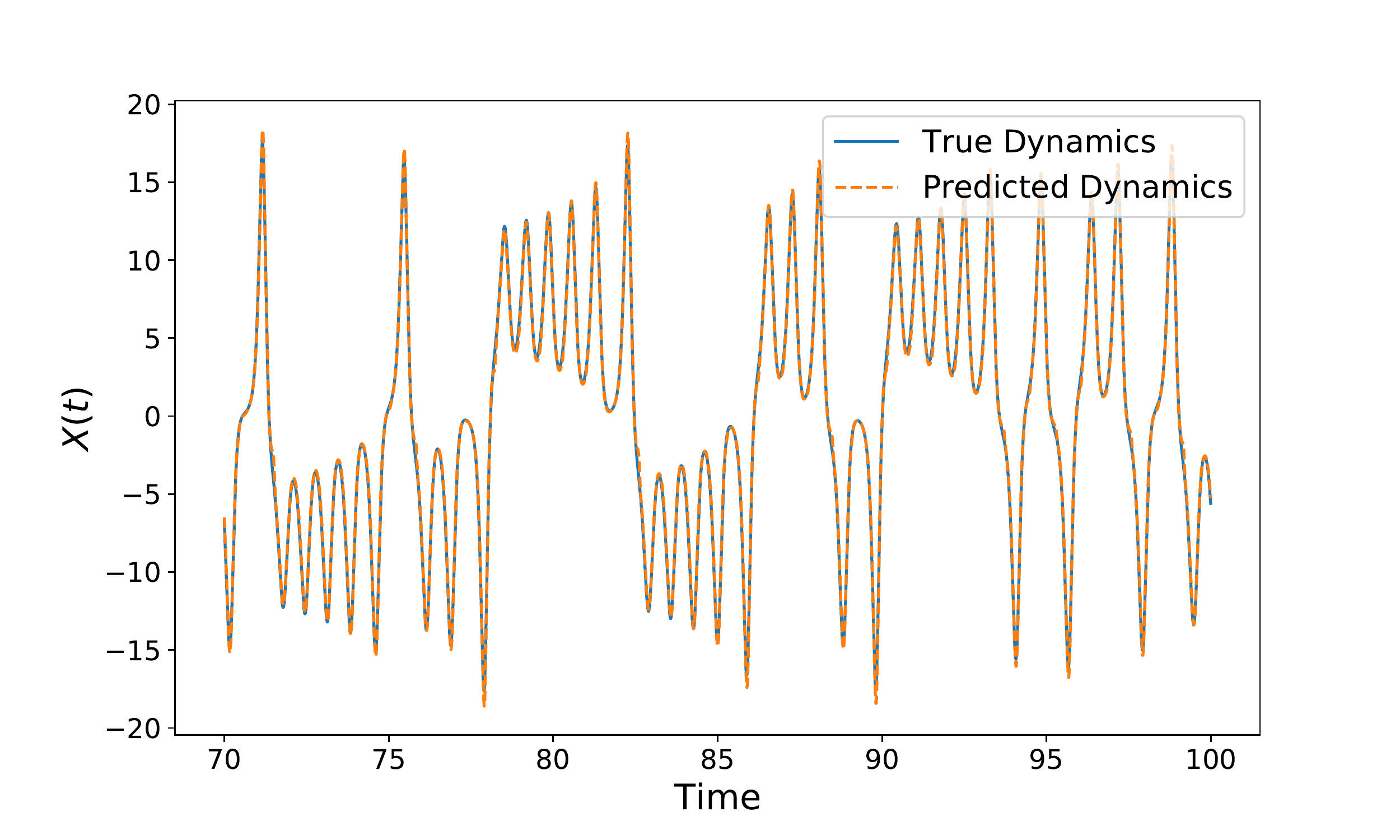}
\caption{ For the  $x$-coordinate of the Lorenz system
\eqref{Eq:5}, the figure shows the predictions in the test set from
the best NN model (dashed orange line) compared with those from
numerical integration of the actual dynamics (continuous blue line). The
learning rate is $\eta=10^{-6}$, while Table $\ref{T2}$ shows the NN architecture.}
\label{Fig-2}
\end{figure}

We trained multiple NN models simultaneously by varying the number of hidden layers, number of nodes in a layer, learning rate and other \textit{hyperparameters}. The model which has the minimum RMSE on the test set is the best model. We have implemented our NN model \footnote{We have used a python package called TensorFlow for performing the numerical computations.} in two different time-series data corresponding to two different systems. The first dataset (Dataset-I) corresponds to the Lorenz system, whose governing equations \cite{Lorenz}  are 
\begin{equation}
\dot{x} = \sigma(y - x), \hspace{0.5 cm}
\dot{y} = x (\rho - z)  - y,  \hspace{0.5 cm}
\dot{z} = xy - \beta z,
\label{Eq:5}
\end{equation}
with $\rho$, $\sigma$ and $\beta$ being the parameters of the system. For values $\sigma = 10$ , $\rho = 28$ , $\beta = 8/3$, the system exhibits chaotic behavior \cite{Lorenz}. The data were obtained by numerically integrating the Lorenz equations by fourth order Runge-Kutta method with parameters in the chaotic regime. We consider the $x$ data to be our input data. After training the NN model, we plot in Fig. \ref{Fig-2}  the predicted evolution of the state $x$ versus time. For comparison, we also plot the actual dynamics in this figure. Clearly, we observe that the yellow dashed line (predicted dynamics) approximates perfectly the blue solid line (actual dynamics). From our analysis, the model was found to be giving 0.38 \% RMSE error in the test set. The neural network model knows nothing a priori about the governing equations of the data, and yet, it is able to learn very well from the training data and predict the data in test set i.e., future motion. Recently, advanced algorithms similar to NN like Reservoir computing  have shown great success in predicting the long term evolution of chaotic systems \cite{RC}. 

\begin{figure}[h!]

\centering
\includegraphics[width=10 cm, height=5.5 cm]{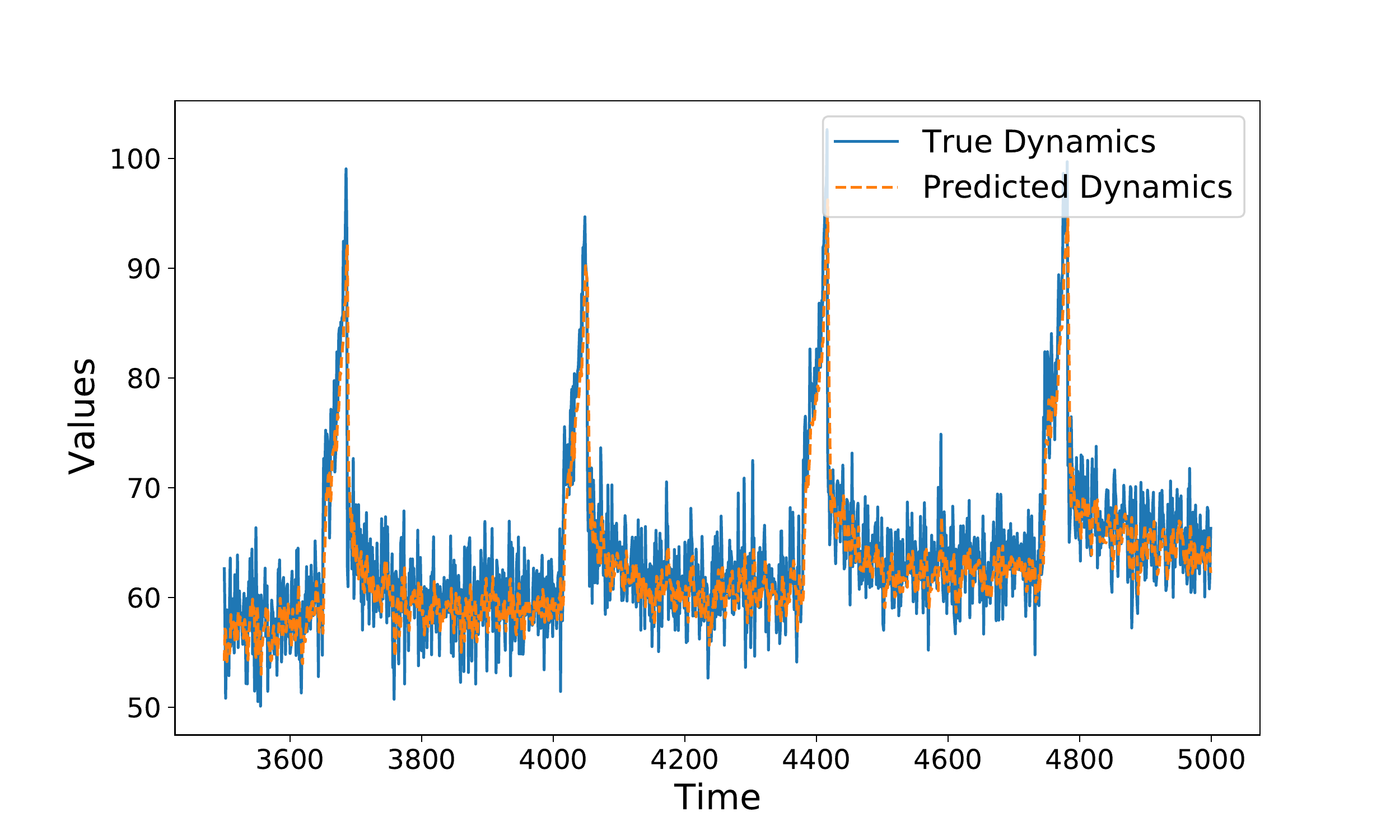}
\caption{The best model of standard  neural network in the Test Set for a generated stochastic time-series data. The solid blue line is the actual dynamics and the  dashed orange line is the predicted dynamics from the NN model. The
learning rate is $\eta=10^{-6}$. NN architecture is shown in Table $\ref{T2}$.}\label{Fig-3}
\end{figure}

The second dataset (Dataset-II) is a noisy time-series data depicting a more practical scenario. We generate these data by adding random numbers (noise) to a periodic time-series data with the average increasing in time. The results are shown in Fig. \ref{Fig-3} with the best model having a 3.88 \% error. We observe from the figure that the NN model has successfully captured the trend, seasonality, and noisy areas of the actual time-series data. Although we do not have any dynamical model for this time-series data, the NN model is able to predict the dynamics sufficiently well. This feature of learning from a finite amount of training data and predicting quite accurately  the data in the test set makes this ML technique quite a powerful and useful tool. All of the aforementioned results and the architecture of the best NN models are summarized in Table \ref{T2}. 

\begin{table}
\begin{center}

\begin{tabular}{|c|c|c|c|c| } 
  \hline
Dataset & Best NN Model & Learning Algorithm & RMSE \\ 
 \hline
Dataset-I  & 20-100-10-1 & GD & 0.38 \% \\ 
Dataset-II  & 20-128-64-64-1 & GD & 3.88 \%\\ 
 \hline
\end{tabular}
\end{center}
\caption{Description of the NN models for the two data sets. The NN architecture is written in the order of the number of nodes in each layer. GD means gradient descent. We have used a modified version of GD called stochastic gradient descent with momentum 0.9 \cite{DLB} to perform our numerical computations.}\label{T2}
\end{table}

\section{Sparse Identification of Nonlinear Dynamical Systems}

The idea of SINDy \cite{SINDy} (Sparse Identification of Nonlinear
Dynamical Systems) is to obtain the governing equations of a nonlinear
dynamical system with $d$ degrees of freedom from the time-series data
of the dynamical state $\bm{x}$ defined as, 
\begin{equation}
\bm{x} \equiv \begin{bmatrix}x_{1} & x_{2} & \ldots & x_{d}\end{bmatrix}.
\label{Eq:6} 
\end{equation}
 Specifically, the objective here is to
obtain a dynamical equation of the form
\begin{equation}
    \dv{\bm{x}(t)}{t} =  \bm{f}(\bm{x}(t)).
    \label{Eq:7}
\end{equation}

Here, the function $\bm{f} \equiv \begin{bmatrix}f_{1} ~ f_{2} ~ \ldots ~ f_{d}\end{bmatrix}$
determines the dynamical evolution of $\bm{x}(t)$. Now, it is a fact
that for most known dynamical systems, the governing equations contain
only a few functions among all possible functions. To understand this statement, consider for
example the Duffing oscillator, which has only linear and cubic terms in
its equations of motion: 
\begin{equation}
    \dot{x} = y, \hspace{0.5cm}
    \dot{y} = - x - \beta x^{3}. 
    \label{Eq:8}
\end{equation}

In matrix form, the above dynamics may be rewritten as
\begin{equation}
\underbrace{\begin{bmatrix}
\dot{x} &
\dot{y}
\end{bmatrix}}_{\bm{\dot{x}}}
= \underbrace{\begin{bmatrix}
1 & x & y & x^2 & xy & y^2 & x^3 & x^{2}y & x y^{2} & y^{3} 
\end{bmatrix}}_{\bm{\Theta(\bm{x})}}
\underbrace{
\begin{bmatrix}
0 & 0 \\
0 & -1 \\
1 & 0 \\
0 & 0 \\
0 & 0 \\
0 & 0 \\
0 & -\beta \\
0 & 0 \\
0 & 0 \\
\underbrace{0}_{w_{1}} & \underbrace{0}_{w_{2}} \\
\end{bmatrix}}_{W},
\label{Eq:9}
\end{equation}
where $\bm{\Theta(\bm{x})}$ is the library of polynomial functions in
($x, y$) up to third order. Clearly, the weight matrix $W \equiv [w_1 ~ w_2]$ for the
Duffing oscillator is sparse (most of its elements are 0). Thus we assume that the equations of motion for any system involve only a few functions among all possible elementary functions represented by the library:
\begin{equation}
\bm{\Theta(\bm{x})} = \begin{bmatrix}
\theta_{1}(\bm{x}) & \theta_{2}(\bm{x}) & \ldots & \theta_{p}(\bm{x}) 
\end{bmatrix}
\label{Eq:10}
\end{equation}
where  $p$ is the total number of candidate functions and $\theta_a$ with $a = 1 ,\ldots, p$ denotes nonlinear functions in $\bm{x}$. Note that $\theta_{a}(\bm{x}) \equiv \theta_{a}(x_1, x_2, x_3, \ldots ,x_d)$.  One may choose the set of functions $\theta_a(\bm{x})$ according to the system of interest.

Each column equation in \eqref{Eq:7} represents the time evolution of a particular component of $\bm{x}$ (such as $\dot{x}$ and $\dot{y}$ for Duffing oscillator \eqref{Eq:8}). This may be written as

\begin{equation}
 \dv{x_k(t)}{t} = \bm{\Theta}(\bm{x})w_{k} \hspace{0.3 cm}  \text{for} \, k = 1, \ldots, d
\label{Eq:11}
\end{equation}

where $w_k$ corresponds to the weight vector consisting of the weights for all the $p$ functions in $\bm{\Theta}(\bm{x})$. 
We compute $\dv{x_k(t)}{t}$ and $\bm{\Theta}(\bm{x})$ based on the given time-series data. The time-derivative of the
time-series data may be obtained either from
measurements or by using suitable numerical differentiation techniques. To determine $w_{k}$, we have to construct a loss function in such a way that
the weight vector $w_k$ is sparse.

This is a supervised learning framework, where $\dot{x}_{k}$ is the actual output data $y_{i}$ and $\Theta(\bm{x})w_{k}$ is
the predicted value of the output from the SINDy model, similar to $f(x_{i},W)$ in the NN model. In both
cases, we learn the parameter set through regression. The loss function which results in a sparse parameter set has the form
\begin{equation}
L_{1}(W) = \frac{1}{N}\sum_{i = 1}^{N} (y_{i} - f(x_{i},W))^{2} +
\lambda \sum_{j = 1}^{M} \vert w_{j} \vert,
\label{Eq:12}
\end{equation}
where $N$ is the total number of data points, $M$ is the total number of parameters and $\lambda$ is a regularization constant. We have discussed intuitively in the appendix why such a loss function is appropriate for learning sparse parameter set. 

Once the numerical values of the elements of the weight vector are determined, the governing equation for each column may be constructed as $\dot{x}_k = \Theta(\bm{x})w_{k} \, \text{for} \, \, k = 1,\ldots, d $ as in Eq. \eqref{Eq:9}.

\subsection{Results and Discussions}

In this section, we discuss results obtained on application of the SINDy
method to two representative nonlinear dynamical systems. The
time-series data required for our
purpose are obtained by employing the usual fourth-order Runge-Kutta
method to integrate numerically the defining equations of motion.\\ 

\textit{The Duffing-Van der Pol Oscillator}\\

\begin{figure}[h!]
\centering
\includegraphics[width= 9 cm, height=6 cm]{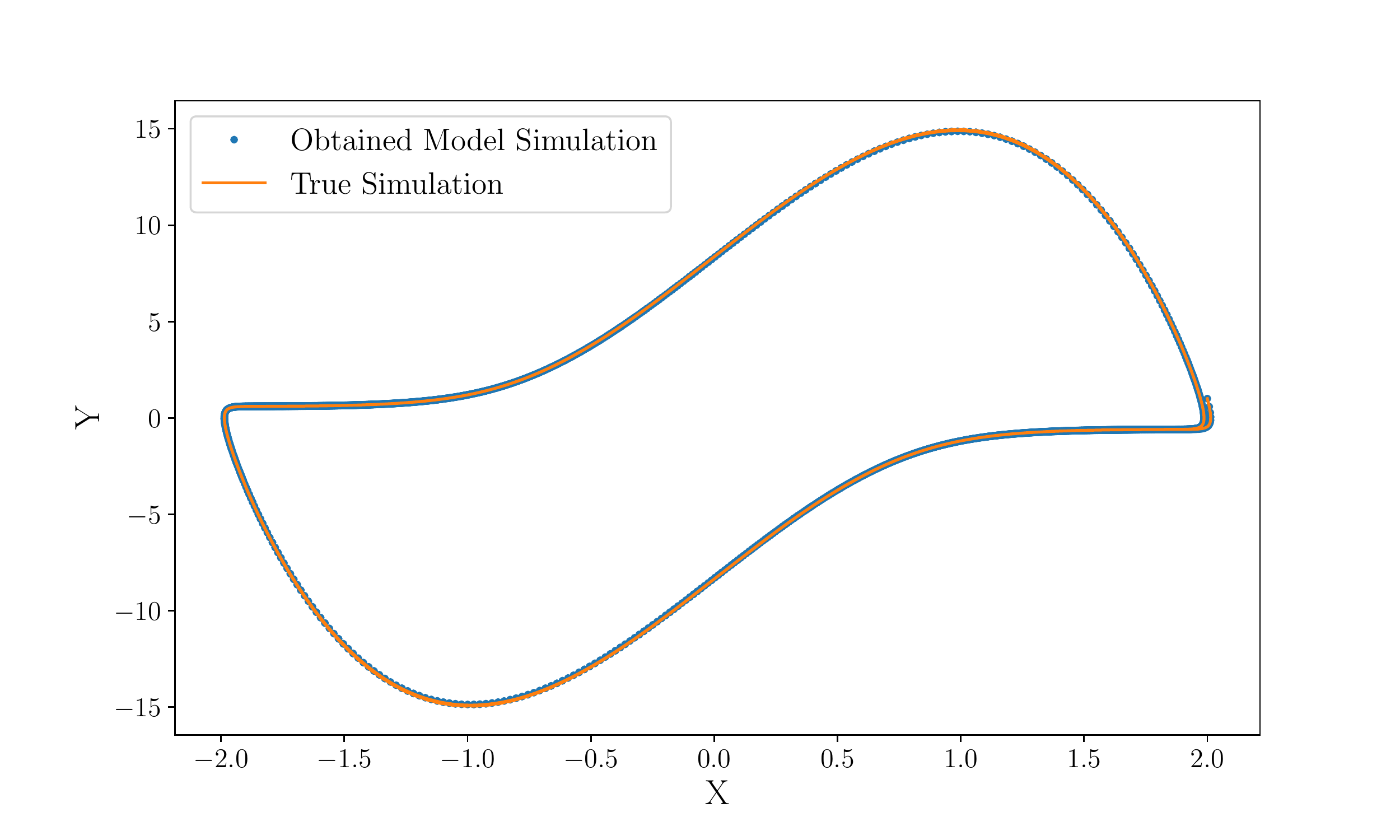}
\caption{Comparison of the dynamical behavior of the Duffing-Van der Pol
oscillator given by Eq. \eqref{Eq:14} (denoted by the orange
continuous line), and the SINDy model given by Eq. \eqref{Eq:15} (denoted by blue dots). }\label{Fig-4}
\end{figure} 

The Duffing-Van der Pol oscillator \cite{VDP} is a particular nonlinear
oscillator that has been studied extensively over the years due to its rich bifurcation behavior and limit-cycle dynamics. The governing dynamics of the oscillator is given by
\begin{equation}
    \ddot{x} - \mu (1 - x^2)\dot{x} + x + \beta x^3 = 0,
    \label{Eq:13}
\end{equation}
where $\mu$ and $\beta$ are real constants representing dynamical parameters. The above dynamics may be
written down as two coupled first-order differential equations:
\begin{equation}
    \dot{x} = y , \hspace{0.5 cm}
    \dot{y} = \mu (1 - x^2)y -x -\beta x^3.
    \label{Eq:14}
\end{equation}
In our analysis, we have used $\mu = 10$, $\beta = 2$. From the SINDy algorithm, we obtain the dynamics: \footnote{We use a python package called pySINDy for performing the
numerical computations.}
\begin{equation}
    \dot{x} = 0.998 y, \hspace{0.5 cm}
    \dot{y} = -1.002 x + 9.914 y -1.995 x^3 -9.914 x^2 y.
    \label{Eq:15}
\end{equation}
We thus see that with the SINDy algorithm, the dynamical
 parameters have been determined within $0.47$ \% of their actual
 values. Moreover, from Fig. \ref{Fig-4}, we may observe that the predicted dynamics
 captures quite successfully the actual behavior.\\
 
\textit{The R\"ossler attractor}\\

Next, we check how the SINDy method performs in the context of a
nonlinear dynamical system displaying a so-called chaotic attractor. We
choose the R\" ossler attractor \cite{Rossler} as our system of study. The governing equations are

\begin{equation}
\dot{x} = -y -z, \hspace{0.5 cm}
\dot{y} = x + a y, \hspace{0.5 cm}
\dot{z} = b + z (x - c), 
\label{Eq:16}
\end{equation}
where the real constants $a,~b,~c$ are the dynamical parameters of the
system. In our analysis, we have chosen the parameter values $a =
0.2,~b = 0.2,~c = 5.7$ for which the R\"ossler system is known to show
chaotic behavior \cite{Rossler}. The SINDy algorithm predicts the
dynamics as 
\begin{eqnarray}
 \nonumber \dot{x} &=& -1.000 y - 1.000 z, \\
\dot{y} &=&1.000 x + 0.200 y,\\ \nonumber
\dot{z} &=&  0.200 - 5.697 z + 1.000 x z.
\label{Eq:17}
\end{eqnarray}

\begin{figure}[h!]
\centering
\includegraphics[width=10 cm, height=6cm]{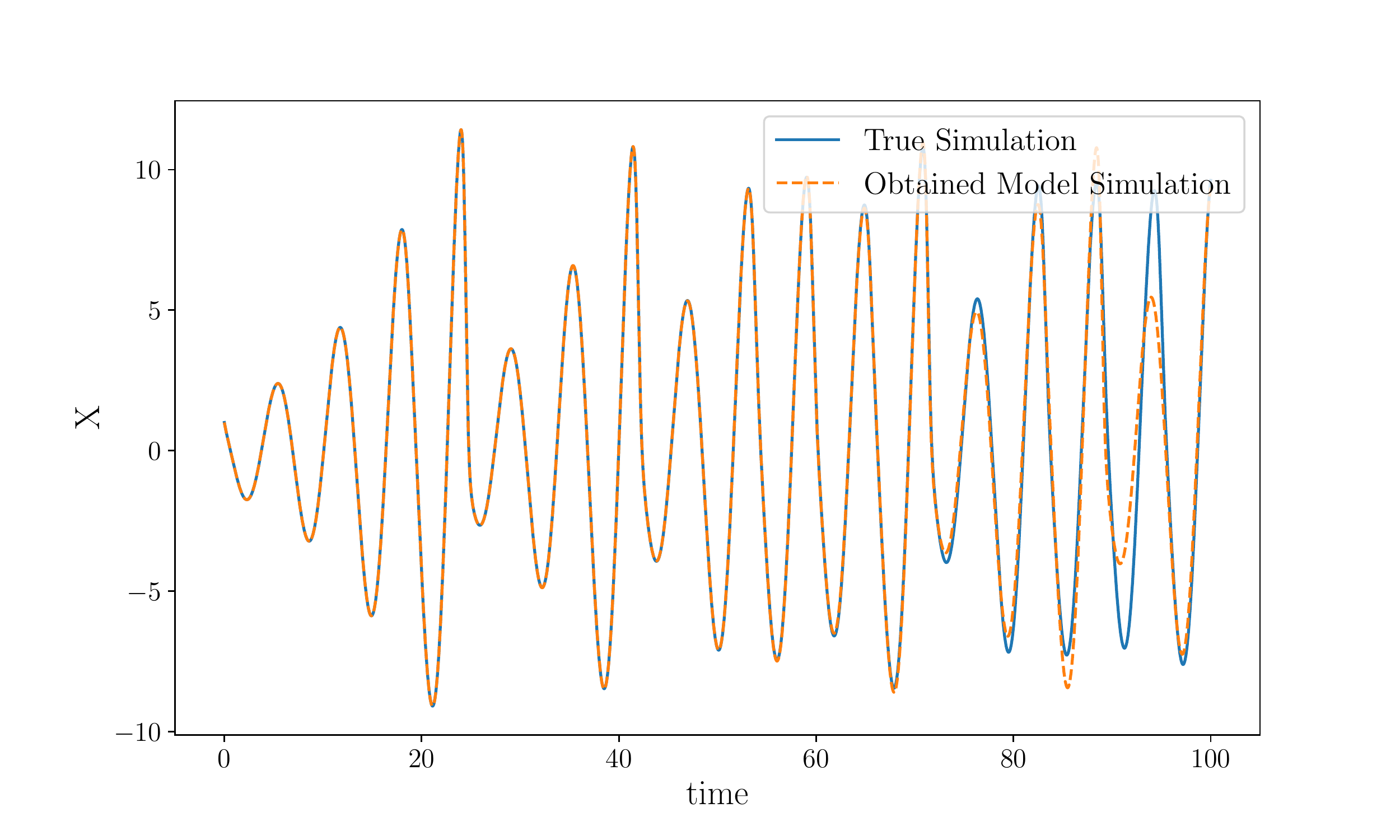}
\caption{Comparison of $X$-coordinate time-series data for R\" ossler attractor (blue solid) and the SINDy model (yellow dashed).}\label{Fig-5}
\end{figure}

\begin{figure}[h!]
\centering
\includegraphics[width=10 cm, height= 5cm]{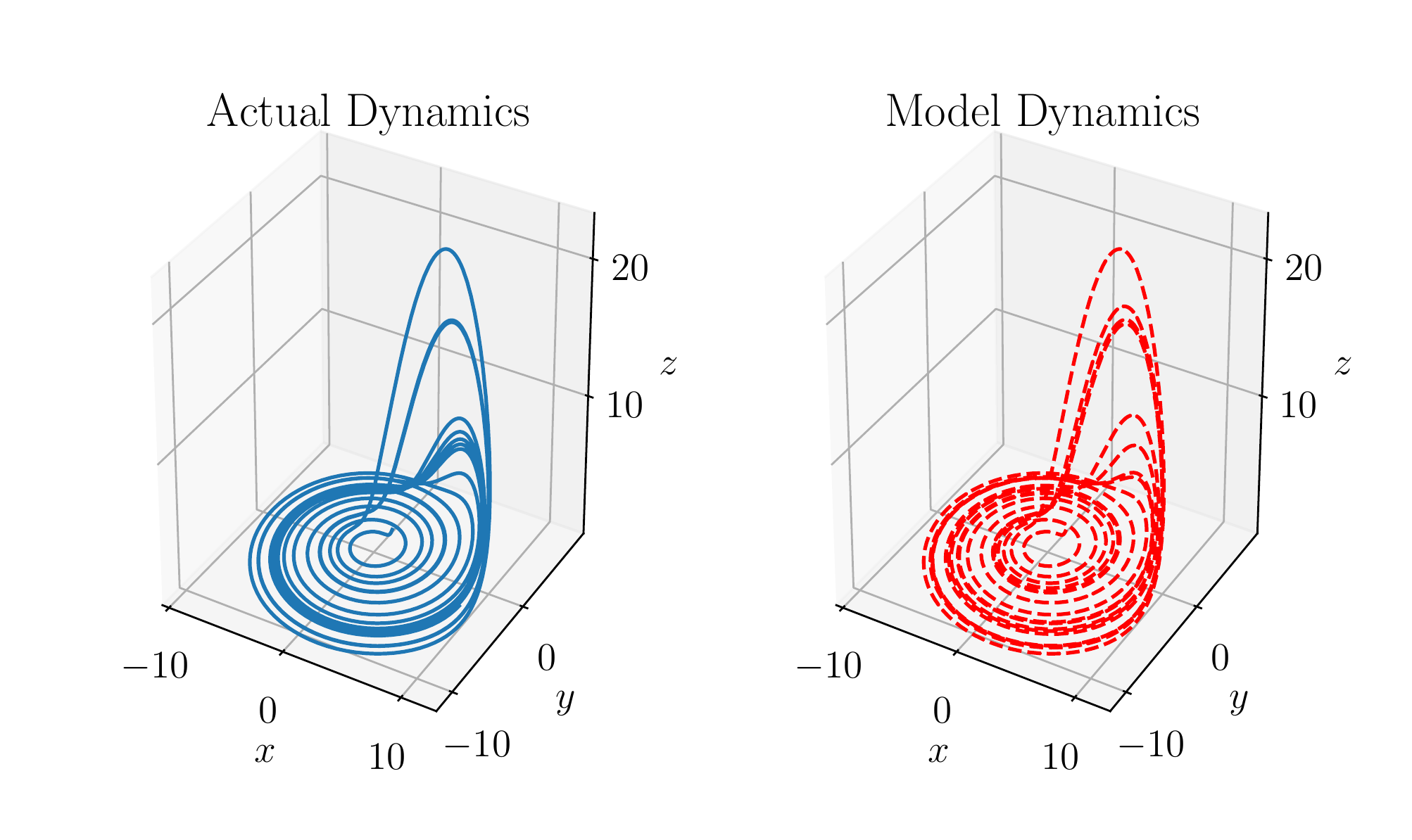}
\caption{Comparison of the actual trajectory of the R\"ossler attractor
(left) and the trajectory of the model obtained by SINDy (right). }\label{Fig-6}
\end{figure}

Chaotic systems are very sensitive to initial conditions: minute
perturbations  lead to very  different dynamics \cite{SH}. The dynamical
parameter values determined by SINDy differ from their actual values by
0.05 \%. This is the reason why at long times, the model trajectory
deviates slightly from the true trajectory, as shown in Fig.
\ref{Fig-5}. Nevertheless, from Fig. \ref{Fig-6}, it is evident that the
model captures qualitatively the dynamics of the R\"ossler attractor.
Namely, had we not labeled the two plots in  Fig. \ref{Fig-6}, one would not have been able to make out as to which one corresponds to the original model and which one to the model predicted by SINDy. 

The above examples serve to establish the fact that the SINDy algorithm
can capture faithfully the actual behavior of two different nonlinear
systems, and thus may prove helpful in modelling dynamics from
given time-series data. The power of this method may be appreciated from
an ML study pursued for a model system, namely, the fluid flow example
of vortex shedding behind an obstacle. This problem took
almost three decades for scientists to unveil its underlying dynamical
equations, while based on time-series data, the dynamics could
be predicted by SINDy within a few minutes \cite{SINDy}!
\section*{Conclusions}

In conclusion, we have demonstrated using paradigmatic nonlinear systems
the use of two different machine learning framework in predicting and
extracting governing  dynamical equations from time-series data. We have
shown that neural network models can prove powerful in predicting the
future evolution of dynamical systems from given data, in cases for
which the exact functional form of the governing equations is not known.
The SINDy method that we have discussed not only identifies the
nonlinearities of the dynamics, but also determines the dynamical
parameters with high precision, and the obtained models capture efficiently the nonlinear behavior of the original dynamical system. 
\section*{Acknowledgement}
Sayan Roy acknowledges DST-INSPIRE, Government of India for providing
him with a scholarship. Debanjan Rana acknowledges KVPY,  Government of
India for providing him with a scholarship. The authors are grateful to
Shamik Gupta for critically reading the manuscript and for insightful
comments on its content. SR also acknowledges Debraj Das for help with figures.

\section*{Appendix: Regularization}

\begin{figure}[h!]
\centering
\includegraphics[width= 9 cm, height= 12cm]{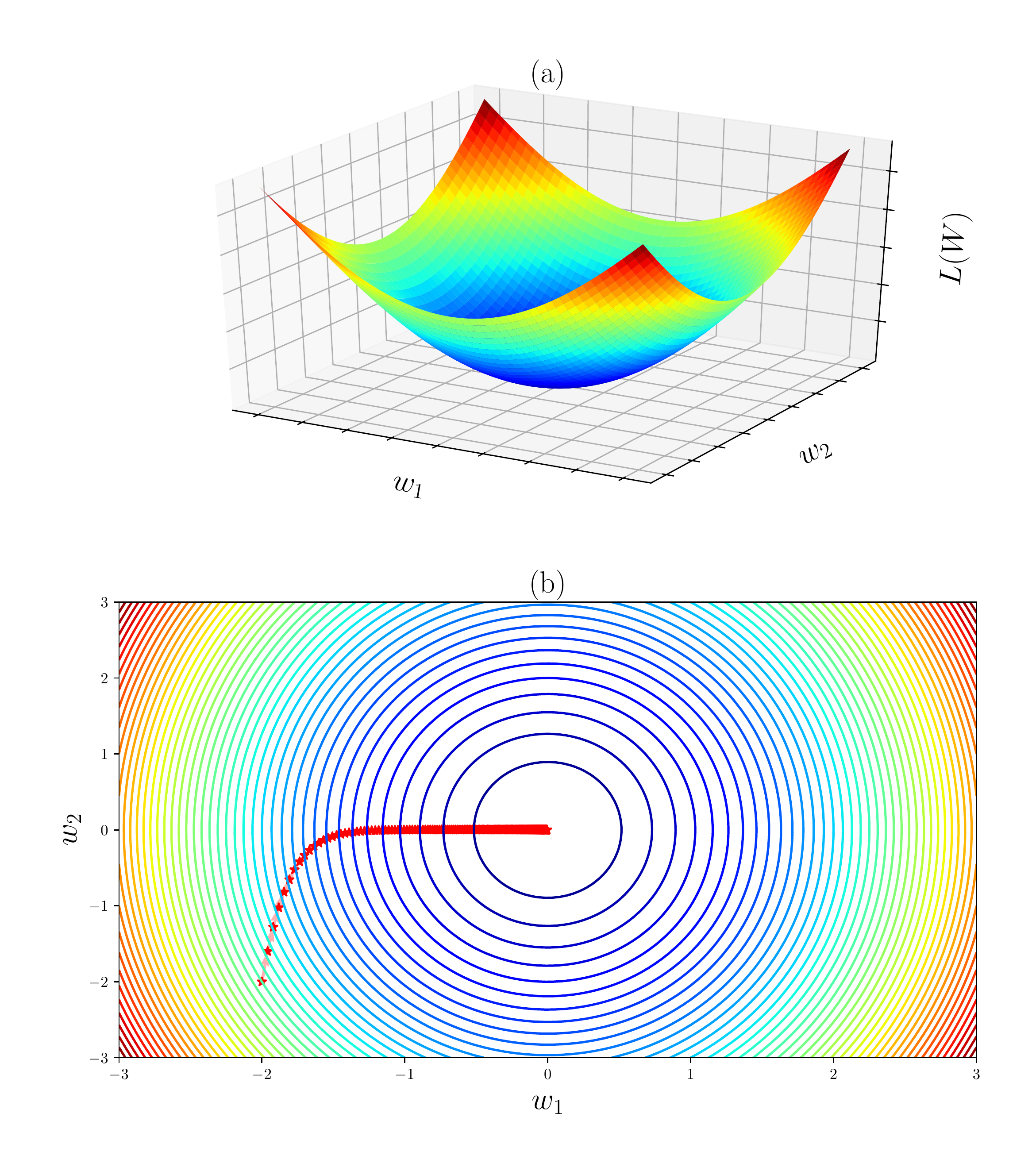}
\caption{Referring to the discussion in the appendix, (a) is the schematic representation of the loss function
$L(W)$, with $w1$ and $w2$ being the two parameters. At
the center of the depicted surface lies the minimum of the loss function. (b) represents the
projection of $L(W)$ onto the $w1-w2$ plane. Every closed contour
corresponds to a given value of $L(W)$. The red dots represent
schematically how the parameters learn about the minimum via gradient descent.}\label{Fig-7}
\end{figure}

In this appendix, we discuss in the context of SINDy algorithm
introduced in Section 3 how one may construct a loss function that
results in a sparse parameter set (most of the parameter values being
0). For simplicity, we will take the number of parameters to be two,
though the discussion here is easily generalizable to the case of a
larger number of parameters. As already mentioned in Section 2, during
the learning process, the loss function has to be minimized with respect
to the parameters $W$. While Fig. \ref{Fig-7}(a) shows the loss function
in the two-dimensional parameter space $\{w1,w2\}$, its projection onto
the $w1-w2$ plane is represented by a closed contour several of which
are shown in Fig. \ref{Fig-7}(b).

 Each contour corresponds to a
different value of the loss function. The outer contours have a larger
value of the loss function with respect to the inner contours.  The goal
of optimization is to learn the parameter values that result in the loss
function having its minimum value, with the latter corresponding to the
center of the contours. Note that the center does not necessarily
correspond to several parameters having zero value. Such an optimization
might also ``overfit" the model in the training data. Overfitting
implies that the model would not predict efficiently for new examples,
thereby severely limiting its applicability. 

\begin{figure}[h!]
\centering
\includegraphics[width=10 cm, height=6 cm]{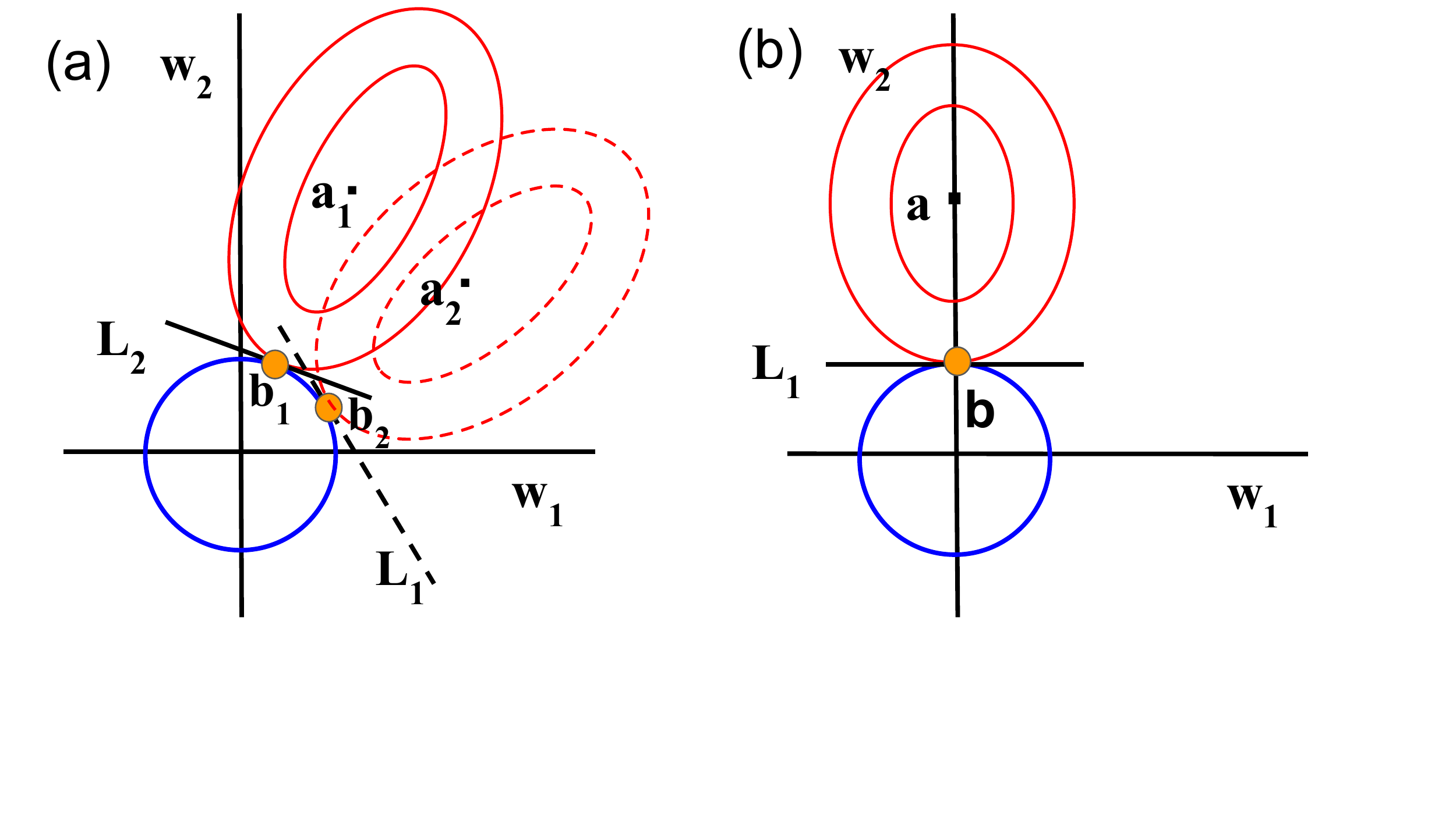}
\caption{Referring to the discussion in the appendix, (a) represents two cases for which the tangent line is not parallel
to either the $x$ or the $y$ axis. The points $b_{1}$ and $b_{2}$ yield the minimum
value of loss function $L_{2}(W)$. They are not sparse. The case of
learned parameters to be sparse is shown in (b). The figures are only schematic. }\label{Fig-8}
\end{figure}

 In order to circumvent the problem of overfitting, we apply regularization. The idea is to learn the parameters for which the error is not precisely minimum (Point (a) in Fig.  \ref{Fig-8}(b)), but has a value corresponding to a nearby contour (Point (b) in Fig. \ref{Fig-8}(b)). The most common is the L2 regularization, defined as 
\begin{equation}
L_{2}(W) \equiv \frac{1}{N}\sum_{i = 1}^{N} (y_{i} - f(x_{i},W))^{2} +
\lambda \sum_{j = 1}^{M} (w_{j})^{2},
\label{Eq:18}
\end{equation}
where $M$ is the total number of parameters and $\lambda$ is the
regularization constant. For $M =2$, Eq. \eqref{Eq:18} is the
mathematical statement of minimizing the loss function $L(w)$ (Eq.
\eqref{Eq:3}) subject to the constraint  $w_{1}^{2} + w_{2}^{2} \leq t$,
where $t$ is a positive constant \cite{GS}. Indeed, it is easy to see
that values of $w$ that satisfy both equation Eq. \eqref{Eq:3} and the constraint would
satisfy Eq. \eqref{Eq:18} for any $\lambda$.  Geometrically, the
constraint represents the area inside the circle $w_{1}^{2} + w_{2}^{2}
= t$, see Fig. \ref{Fig-8}. The parameters that satisfy both the
conditions would correspond to the point at which both the contour and
the circle have a common tangent (point b in Fig. \ref{Fig-8}(b)). An
outer contour than the  one corresponding to this point will satisfy the
constraint condition but will have a larger value of the loss function, while an inner contour will not satisfy the constraint condition. For the parameter set to be sparse, the tangent line has to be parallel to the $x$-axis (shown in Fig. \ref{Fig-8}(b)) or the $y$-axis. Any other tangent line (shown in Fig. \ref{Fig-8}(a)) will result in a non-sparse parameter set. It is then evident the parameters learned this way will have a rather low likelihood of being sparse.

\begin{figure}[h!]
\centering
\includegraphics[width=10 cm, height=6 cm]{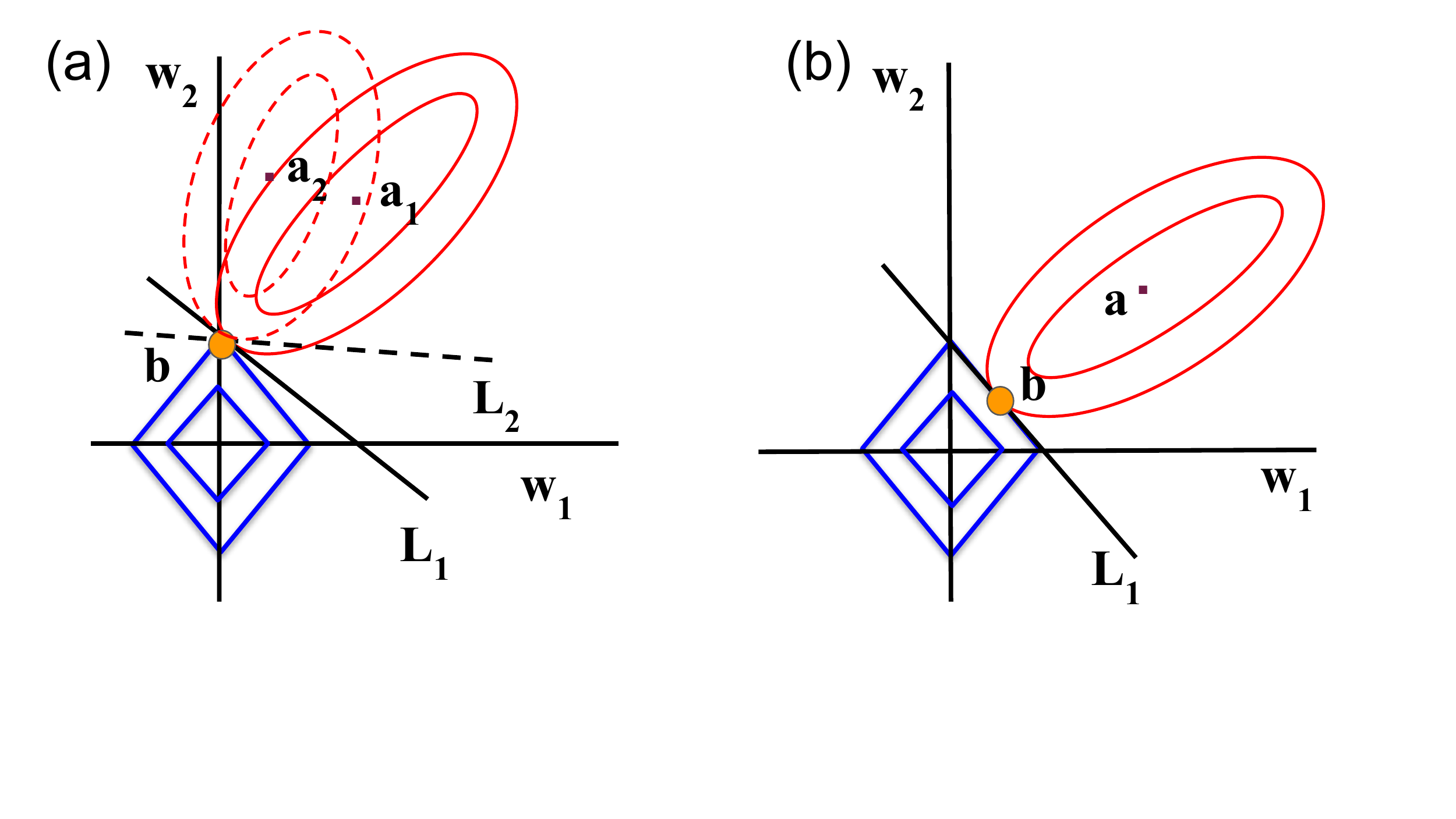}
\caption{Referring to the discussion in the appendix, (a) represents the case in which the tangent is not parallel to
the edges of the diamond. In such cases, the corners of the diamond
represent the minimum value of $L_{1}(W)$, leading to sparsity in
learned parameters. One of the cases of non-sparsity is shown in (b).
The figures are only schematic.  }\label{Fig-9}
\end{figure}

Another type of regularization is L1 regularization, where the loss
function is defined by Eq. \eqref{Eq:12}. Here, for $M =2$, Eq.
\eqref{Eq:12} is derived from the constraint condition $\vert w_{1}
\vert + \vert w_{2} \vert \leq \text{constant}$ \cite{GS}. This
condition corresponds to the area covered by the diamond shown in Fig.
\ref{Fig-9}. Here, the possible case for not having a sparse parameter
set corresponds to the aforementioned tangent line being parallel to any
of the edges of the diamond, as shown in Fig. \ref{Fig-9}(b). Any other
tangent line such as the one shown in Fig. \ref{Fig-9}(a) will intersect
the diamond at the $x$-axis or the $y$-axis corresponding to a minimum
value of the loss function. In this case, the learned parameters will
have a very high likelihood of being sparse.

On the basis of the above discussions,
we intuitively understand that a curve with corners such as a diamond
has a higher probability of generating a sparse parameter set than a
smooth curve such as a circle. For L1 regularization, one would have
more corners with the increase in the number of parameters and
consequently a sparser parameter set. A rigorous mathematical
demonstration of convergence of L1 regularization to a sparse solution
is given in \cite{SL}.



\begin{thebibliography}{} 
\bibitem{CM}
A. Ghosh, \textit{The little known story of F = ma and beyond}, Resonance 14, 1153 (2009).

\bibitem{SH}
 S. H. Strogatz, \textit{Nonlinear Dynamics And Chaos: With Applications To Physics, Biology,
Chemistry, And Engineering}, Westview Press, Boulder (2014).

\bibitem{DLB}
I. Goodfellow, Y. Bengio and A. Courville, \textit{Deep Learning}, MIT Press (2016).

\bibitem{TM}
T. M. Mitchell, \textit{Machine Learning}, McGraw-Hill, New York (1997).

\bibitem{SINDy} 
S. L. Brunton, J. L. Proctor and J. N. Kutz, \textit{Discovering governing equations from data by sparse identification of nonlinear dynamical systems}, Proceedings of the National Academy of Sciences, 113(15), 3932-3937 (2016).
 
 \bibitem{Lorenz}
E. N. Lorenz, \textit{Deterministic nonperiodic flow}, Journal of the Atmospheric Sciences, 20(2), 130–141 (1963).

\bibitem{RC}
J. Pathak, B. Hunt, M. Girvan, Z. Lu, and E. Ott, \textit{
Model-Free Prediction of Large Spatiotemporally Chaotic Systems from Data: A Reservoir Computing Approach}, Phys. Rev. Lett. 120, 024102 (2018).

\bibitem{VDP}
B. Van der Pol, \textit{On relaxation-oscillations}, The London, Edinburgh and Dublin Phil. Mag. \& J. of Sci., 2(7), 978–992 (1926).


\bibitem{Rossler}
O. E. R\" ossler, \textit{An Equation for Continuous Chaos}, Physics Letters, 57A(5), 397–398(1976).

\bibitem{GS}
H. Goldstein, \textit{Classical Mechanics}, Addison-Wesley Publishing Company (1980).


\bibitem{SL}
J. H. Friedman, R. Tibshirani and T. Hastie, \textit{The Elements of Statistical Learning: Data Mining, Inference, and Prediction}, New York : Springer, (2009).


\end{thebibliography}
\end{document}